\documentclass[12pt,preprint]{aastex}
\def\ltsima{$\; \buildrel < \over \sim \;$}
\def\simlt{\lower.5ex\hbox{\ltsima}}
\def\gtsima{$\; \buildrel > \over \sim \;$}
\def\simgt{\lower.5ex\hbox{\gtsima}}
\def\eps@scaling{.95}
\def\plotone#1{\centering \leavevmode
   \epsfxsize=\eps@scaling\columnwidth \epsfbox{#1}}

\shorttitle{Merger Origin of the Thick Disk}
\shortauthors{Wyse et al.}
\received{}

\begin{document}
\title{Further Evidence for a Merger Origin for the Thick Disk: Galactic Stars Along Lines-of-sight to Dwarf Spheroidal Galaxies}

\author{Rosemary F.G.~Wyse\altaffilmark{1},
Gerard~Gilmore\altaffilmark{2}, John E.~Norris\altaffilmark{3}, M.~I.~Wilkinson\altaffilmark{2},
Jan T.~Kleyna\altaffilmark{4}, A.~Koch\altaffilmark{5}, N.W.~Evans\altaffilmark{2}, E.~K.~Grebel\altaffilmark{5}
}
\altaffiltext{1}{The Johns Hopkins University, Dept.~of Physics and Astronomy, Baltimore, MD 21218; wyse@pha.jhu.edu}
\altaffiltext{2}{Institute of Astronomy, Cambridge University, Madingley Road, Cambridge CB3 0HA, UK; gil@ast.cam.ac.uk}
\altaffiltext{3}{Research School of Astronomy \& Astrophysics, The Australian National  University, Mount Stromlo Observatory, Cotter Road, Weston, ACT 2611; jen@mso.anu.edu.au}
\altaffiltext{4}{Institute for Astronomy, University of Hawaii, 260 Woodlawn Drive, Honolulu, HI 96822; kleyna@ifa.hawaii.edu}
\altaffiltext{5}{Astronomical Institute of the University of Basel, Department of Physics \& Astronomy, Venusstr. 7, CH-4102 Binningen, Switzerland; koch@astro.unibas.ch, grebel@astro.unibas.ch}

\begin{abstract}

The history of the Milky Way Galaxy is written in the properties of
its stellar populations.  Here we analyse stars observed as part of
surveys of local dwarf spheroidal galaxies, but which from their
kinematics are highly probable to be non-members. The selection
function -- designed to target metal-poor giants in the dwarf
galaxies, at distances of $\sim 100$~kpc -- includes F-M dwarfs in the
Milky Way, at distances of up to several kpc. 
The
stars whose motions are analysed here lie in the cardinal directions
of Galactic longitude $\ell \sim 270^\circ$ and $\ell \sim 90^\circ$,
where the radial velocity is sensitive to the orbital rotational
velocity.  We demonstrate that the faint F/G stars contain a
significant population with $V_\phi \sim 100$~km/s, similar to that
found by a targeted, but limited in areal coverage, survey of
thick-disk/halo stars by Gilmore, Wyse \& Norris (2002). This value of
mean orbital rotation does not match either the canonical thick disk
or the stellar halo. We argue that this population, detected at both $\ell \sim 270^\circ$ and $\ell \sim 90^\circ$, 
has the expected properties of `satellite debris' in the
thick-disk/halo interface, which we interpret as remnants of the
merger that heated a pre-existing thin disk to form the thick disk.

\end{abstract}


\section{Introduction} 

The Milky Way Galaxy, once apparently satisfactorily described by
Population I and Population II, is clearly a very complex system
(Freeman \& Bland-Hawthorn 2002). Substructure in kinematics and in
coordinate space indicates accretion of (formerly) satellite galaxies
(e.g.~the Sagittarius dwarf spheroidal: Ibata, Gilmore \& Irwin 1994; Ibata et al.~1997; Majewski et
al.~2003) and disruption of star clusters (e.g.~Odenkirchen et
al.~2003).  Merger and assimilation of satellite galaxies is an inherent part of the formation of large galaxies, like the Milky Way, in the now-canonical $\Lambda$-Cold-Dark-Matter cosmological models.  Indeed 
thick disks of spiral galaxies are most probably  relics from an early merger of the
evolving thin disk and a smaller satellite galaxy (see Wyse 2005 for a recent discussion of the possibilities), and thus contain 
unique clues about the early stages of disk galaxy formation.  

Models of the formation of thick disks through heating of a thin disk by (dissipationless) mergers
usually invoke a satellite of 10-20\% of the mass of the
(pre-existing) 
stellar disk (e.g.~Velaquez \& White 1999; Walker,
Mihos \& Hernquist 1996).  The outer parts of the satellites are
removed by tidal forces, limited by the relative densities of the
satellite and larger system (essentially a Roche-Jacobi criterion). The
fate of this debris has been largely ignored in the models, but it is
clear that material stripped by tidal forces continues on orbits close
to that of the center of mass of the satellite at the time of removal.
Thus, for typical satellite/subhalo initial orbital angular momentum
that is close to half that of a circular orbit of the same energy
(e.g.~Benson 2005; Zentner et al.~2005) one would expect an azimuthal
streaming velocity of around 100km/s for satellite debris in a system
like the Milky Way with a flat rotation (circular velocity) curve,
with amplitude \footnote{Of course in hierarchical clustering models
the potential well of the Milky Way is not fixed, but observations
constrain the bulk of the merging for normal galaxies to happen very
early (e.g.~Glazebrook et al.~2004; Unavane,
Wyse \& Gilmore 1996; Wyse 2005), and theory predicts an early epoch of massive
mergers (e.g.~Zentner \& Bullock 2003).} $\sim 220$~km/s.
Identification of this satellite debris would provide strong evidence
for a merger history of the thick disk. 

First indications of a population that could be ascribed to debris
from the satellite whose merger caused the thick disk was presented by
Gilmore, Wyse \& Norris (2002; hereafter GWN).  This population was identified in a
survey of faint (apparent V-band magnitude $V \simgt 18$) F/G stars in
two lines of sight at intermediate latitude, at longitude $\ell \sim
270^\circ$. The primary signature was kinematics distinct from those
of the canonical thick disk, and distinct from the stellar halo, with
mean azimuthal streaming velocity $\sim 100$km/s.  Further analysis
(Norris et al.~2006) has derived a characteristic metallicity about a
factor of ten below the canonical thick disk, and more characteristic
of stars in dwarf galaxies.  But is this a pervasive
population, reasonably well-mixed as expected if indeed
it resulted from a merger that took place many Gyr ago?

\section{Stars in Lines-of-Sight to selected Dwarf Galaxies}

A side benefit of the recent effort to obtain detailed internal
kinematics across the face of local dwarf spheroidal (dSph) galaxies
is reasonably large samples of `contaminating' stars in the Milky Way,
along the line-of-sight to the target dwarf galaxy.  
The cardinal longitudes $\ell \sim 270^\circ$ and $\ell \sim
90^\circ$, where the radial velocity is most sensitive to azimuthal
velocity (and hence orbital angular momentum, given a distance), are
of particular importance.  We here analyse the non-member stars from
our large surveys of the Draco dSph ($\ell =86^\circ,\, b=
+35^\circ$), the Ursa Minor dSph ($\ell =104^\circ,\, b= +45^\circ$),
and the Carina dSph ($\ell =260^\circ,\, b= -23^\circ$).  The data
were taken, in order of sample size, from surveys with the VLT/FLAMES,
WHT/WYFFOS and Keck/HIRES, and have typical radial velocity errors of
less than 3km/s.
 
The stars that were observed were selected from color-magnitude
diagrams of each dSph to be giants at the distance of that
galaxy, with a bright magnitude limit of $V \sim 17$ and faint limit
of $V \simlt 20$, with a mild trend, following the giant branch, that
fainter stars are bluer. The surveys of the Draco dSph and the UMi
dSph are described in Wilkinson et al.~(2004), while that of the
Carina dSph is described in Wilkinson et al.~(2006) and in Koch et
al.~(2006).  In general dSph have low velocity dispersions, $\sim
10$~km/s, and our membership probabilities were based on iterative
Gaussian fitting to the radial velocity distributions.  The member,
and non-member, stars for each line-of-sight are shown in the
color-radial velocity plane in Fig.~1.  The non-members are expected
to be predominantly dwarf stars within the Milky Way at typical
distances of a few, to several, kiloparsec.  As may be seen in the
Figure, in the
Draco dSph line-of-sight the member stars are well-separated from the field
stars but for the 
Carina and UMi dSph lines-of-sight we may well have incorrectly assigned a small number of actual
field stars as high-probability galaxy members.  In the analysis below, this
would result in a (small) deficit of stars in the Carina line-of-sight
with azimuthal streaming velocity $V_\phi \sim 0$~km/s and in the UMi
line-of-sight a (small) deficit of stars with $V_\phi \simlt
-150$~km/s.  There is no {\it a priori\/} kinematic selection bias in the
non-member stars.

\section{Azimuthal Streaming Velocity}

The lines-of-sight are ideal  to probe the orbital angular momentum
of the Galactic stars.  Defining $V_{helio}$ to be the heliocentric radial (line-of-sight) velocity of a given star, then with 
$V_{Gal}$ being the Galactocentric radial velocity of a given star, i.e.~the component of the star's Galactocentric
velocity along the line-of sight from the Sun to the star, we have 
$ V_{Gal} = V_{helio} + v_{pec,\odot,(\ell,b)} + 220 {\rm (km/s)}
\sin \ell \cos b. $ The second term here is the component of the
Sun's peculiar motion in the line-of-sight to the star, and we have
assumed that the LSR has a circular orbit of 220~km/s about the
Galactic Center.  We adopted a solar peculiar motion of 15.4~km/s in
the direction of ($\ell,\, b) = 51^\circ, \, 23^\circ$.  $V_{Gal}$ may
then be decomposed as 
$V_{Gal} = \alpha V_r +
\beta V_\phi + \gamma V_\theta$, where the coefficients are functions
of Galactocentric distance plus angular coordinates, and are defined
in equations 3-7 of Morrison, Freeman \& Flynn (1990; hereafter MFF).
With the assumption of zero means in the $r$ and $\theta$ motions, $V_{gal}/\beta$ for a given star is an unbiased estimate of $V_\phi$. 
For example, for a solar distance  of 3~kpc (a typical derived
value below),
$ V_{Gal,Carina\,los} = 0.48 V_r -0.81  V_\phi +0.33 V_\theta, $ 
$ V_{Gal,Draco\,los} = 0.31 V_r +0.80  V_\phi -0.52 V_\theta,$ and 
$V_{Gal,UMi\,los} = 0.49 V_r +0.62  V_\phi -0.61 V_\theta.$

Implementation of these formulae requires distances. 
We utilised two isochrones from Bergbusch \& VandenBerg
(2001), with color-effective temperature relations from 
VandenBerg \& Clem (2003), and  assumed that the
stars are main sequence dwarfs (we discuss possible subgiant
contamination in section~4 below). We adopted the V$-$I, $M_V$
relation corresponding to 12~Gyr old populations with ${\rm
[\alpha/Fe] = +0.3}$ and firstly $ {\rm [Fe/H] = -0.525}$ and secondly
$ {\rm [Fe/H] = -1.525}$.  The first was chosen since it corresponds
to a large fraction of local thick disk stars (Bensby et al.~2005), while 
the second was
chosen since this metallicity corresponds to the stars with
intermediate kinematics from GWN, as well as to the `metal-weak thick disk' (e.g.~MFF; Chiba \& Beers 2000) and to the stellar halo. 
 Individual distances cannot be
taken as reliable, but the overall derived distributions 
should be representative.    The reddenings in the
lines-of-sight of interest are all small, $E(B-V) \simlt 0.05$
(Schlegel et al.~1998), and are ignored in the analysis.  The derived
distances for the vast majority of the stars are less than 5~kpc, with
a typical distance being 2--3~kpc. 
At the faint magnitudes we analyse below,  the redder stars
are most likely to be in the thin disk -- thick disk interface, with
the bluer ones in the thick disk -- halo interface. 

{\it Carina Line-of-Sight, $\ell =260^\circ,\, b= -23^\circ$:}

The line-of-sight to the Carina dSph contains the largest sample of
non-member stars, 619 stars with apparent magnitudes in the range $
17.27 \leq V \leq 20.31$ and with color in the range $0.66 \le V-I \le
2.61$.  The vast majority are most probably normal disk stars. We here
are most interested in the bluer, fainter stars.  We are also
interested in a comparison of the derived $V_\phi$ distributions in
the three dwarf spheroidal lines-of-sight, and this is facilitated
with a similar color range for all three. We therefore pruned the
sample to the 89 stars with $ 19.0 < V \leq 20.0$ and with color in
the range $0.90 < V-I \le 1.1$.  The Carina input sample was derived
from the $V, \, B-V$ distribution, and the $V-I$ selection here was
adopted to compare with the other two lines-of-sight, which utilised
the $V, \, V-I$ distribution.  The histogram of derived individual
estimates of $V_\phi = V_{Gal}/\beta$ for this `G star sample' is
shown in Figure~2, for each of the two isochrones, together with the
distribution in the $V, \, V-I$ plane.  Note that the bin size,
20~km/s, is chosen to provide significant numbers of stars per bin,
and is more than four times the line-of-sight velocity uncertainty.  The
narrow peak at $\sim 200$~km/s, with the shoulder at $\sim 150$~km/s
is due to the thin disk plus canonical thick disk (see Table~1). There
is a rather high tail through $V_\phi \sim 100$~km/s, towards $V_\phi \sim 0$~km/s.  These results are insensitive to the adopted metallicity (compare the solid and dashed histograms in the figure).

The thick disk/halo interface, of most interest to probe satellite
debris, is best probed by the stars with $V-I \leq 0.9$. Figure~3
shows the velocity histogram of these 49 stars (all fainter than
$V=18$) compared with the 1043 stars (also $V \geq 18$, $V-I \leq
0.9$) in the field at $\ell \sim 270, \, b \sim +33$ that formed part
of the sample of GWN.  Given the difference in latitude, and the
resultant difference in projection of $V_\phi$ into the line-of-sight,
the agreement is good.

{\it Draco Line-of-Sight, $\ell =86^\circ,\, b=
+35^\circ$ }:

The 207 non-member stars in this line of sight cover the apparent
magnitude range $17.00 \leq V \le 19.88$, and colors as shown in
Fig.~1.  We selected a `G-star sample' consisting of the 82 stars with
$ 18.25 < V \leq 19.75$ and with color in the range $0.90 < V-I \le
1.1$.  Their derived $V_\phi$ distribution is shown in Fig.~2, again
with the $V, \, V-I$ color-magnitude diagram (the effects of the
selection of stars following the locus of a red giant branch have not
been entirely removed by these cuts).  The histograms are similar to
those in the Carina line-of-sight, except that the narrow peak is at
the canonical thick disk mean $V_\phi \sim 170-180$~km/s.  Again the
results are insensitive to the metallicity of the adopted isochrones.
This agreement with the Carina line-of-sight and the earlier
GWN data is all the more important since this Draco line-of-sight is
at opposite Galactic longitude, $\ell =90^\circ$ as opposed to $\ell =
270^\circ$.

{ \it UMi Line-of-Sight, $\ell =104^\circ,\, b=
+45^\circ$}:

The 105 non-member stars in this line of sight cover the apparent
magnitude range $16.59 \leq V \le 19.86$, with $V-I$ distribution
shown in Fig.~1, 
We again selected a `G-star sample', this time consisting of the 42 stars with
$ 17.75 < V \leq 19.75$ and again with color in the range $0.90 < V-I \le
1.1$.  Their derived $V_\phi$ distribution and CMD are shown in Fig.~2, (again the effects of the initial 
selection of red giant stars in the dSph  have not
been entirely removed by these cuts).  The effects of small number statistics limit the interpretation of the histograms but the distribution is clearly rather broad. 
Once more  the
results are insensitive to the metallicity of the adopted isochrones.
Again, we note that  this UMi  line-of-sight is at  Galactic
longitude $\ell =90^\circ$, as opposed to $\ell = 270^\circ$.

\section{Comparison with Expectations}

A direct comparison of the velocities with predictions is not
trivial due to the non-standard selection function of the samples,
which were chosen for reasons having nothing to do with studying the
Galaxy. However, the narrower color range of the `G-star' samples
alleviates this to some extent (and was chosen with a model comparison
in mind). Assuming uniform sampling across the color-magnitude ranges
selected, each of the `G-star' samples in the Draco and UMi
lines-of-sight should, from the Gilmore star-count model (Gilmore
1981), be a mix of thin disk, thick disk and halo in the ratio $\sim
0.1:1:0.2$.  This is consistent with the derived distances, which
correspond to mean $z-$height above the Galactic plane of $\sim
1.75$~kpc for the [Fe/H] $\sim -0.5$ isochrone, and $\sim 1.3$~kpc for
the [Fe/H] $\sim -1.5$ isochrone.  The stars in the Carina
line-of-sight should be a mix of $\sim 0.4:1:0.2$, again consistent
with our derived mean height above the Galactic plane of 1.5~kpc for the [Fe/H] $ \sim -0.5$ isochrone and 1.1~kpc for the $-1.5$ isochrone. 

The highest statistical significance comes from the sum of all three
lines-of-sight, and the lower right panel of Fig.~2 shows a simple sum
of the three individual histogram, together with a model curve that is
the sum of three Gaussians to
represent the thin and thick disks and halo, with  means and sigmas given in 
Table~1,
in the ratios of 0.1:1:0.25 (as a compromise fit to the main thin disk/thick disk peak at $\sim 190$~km/s).  There is a clear excess, compared to the
model, of stars with $V_\phi \sim 50-150$~km/s.  This excess is robust
to rebinning (the structure around the peak at $\sim 200$~km/s is more
sensitive).  The usual Kolmogorov-Smirnoff test for comparisons of
data and models is most sensitive to regions with most stars, and thus
application to the entire distribution is dominated by the peak at
around $\sim 200$~km/s, where the agreement is reasonable.  We have
instead applied the KS test to a restricted range of $V_\phi$, finding
for $V_\phi < 100$~km/s there is a  99\% probability that the data are from a different distribution than
the model.  
We obtain 
similar excesses at $V_\phi \sim 100$~km/s (but with lower
significance due to small number statistics) from the $\ell \sim
270^\circ $ Carina line-of-sight alone, and from a sum of the $\ell
\sim 90^\circ$ Draco and UMi lines-of-sight.  The addition of a
component with $V_\phi = 100$~km/s and $\sigma_{V\phi} = 40$~km/s, and
with normalisation equal to that of the stellar halo, greatly improves the fit.

One is required to adopt rather extreme kinematics for the canonical
components of the Galaxy to avoid the addition of this extra
component; for example, increasing both the thin disk and thick disk
dispersions and adopting a mean rotational streaming of 30~km/s for
the halo.  Even with this, one cannot obtain a good fit to the
distributions, but rather introduce a significant over-prediction in
the model around $V_\phi \sim 150$~km/s. Vertical gradients in
$<V_\phi>$ for the thick disk have been discussed (e.g.~MFF, Chiba \&
Beers 2000; Allende Prieto et al.~2006) but none predicts as low a
value of $<V_\phi> \sim 100$~km/s at $z < 2$~kpc.

\section{Discussion} 

There are several possible effects that could complicate the
interpretation of the above analysis.  First, imperfect subtraction of the  several strong
sky lines in the observed data  
could have produced spurious radial
velocities at certain values set by the wavelength of the sky line. We
found that the strongest spurious signal
(itself weak) in the cross-correlation peak 
would be expected at $\sim -145$~km/s and
$\sim +240$~km/s, far from our reported signal.

We assumed that all stars were dwarfs when estimating distances.  For
the isochrones we adopted, subgiants with $V-I \sim 0.9$  are
some 2--3 magnitudes brighter in the V-band than are the main sequence
stars of the same color. At the apparent magnitudes of interest,
subgiants would lie at distances of 10--15~kpc. The resultant
deprojections of the line-of-sight velocity into $V_\phi$ would shift
the peak at $\sim 200$~km/s (see Fig.~2) to $\simgt 300$~km/s,
incompatible with all other surveys of the thick disk -- thin disk
interface.  The $V_\phi \sim 200$~km/s peak contains stars of all colors, as
does the velocity range around $\sim 100$~km/s, as may be
inferred from Fig.~1 (remembering the typical $|\beta| \sim 0.8$).  Clearly subgiant contamination must be minor,
in line with the expectations from stellar evolution that this phase
is short-lived compared to the main sequence, and with observed
luminosity functions.

We have adopted isochrones based on our knowledge of chemical
abundances in the thick-disk -- halo interface, but clearly
metallicity determinations for the non-member stars would be extremely
useful, particularly to investigate any possible connection to the `metal-weak thick disk' (e.g.~Chiba \& Beers 2000; MFF). This is being investigated.

\section{Conclusions}

We have shown that faint F/G stars, $V \simgt 18$, at intermediate
latitudes, consistently show a population with mean azimuthal
streaming that corresponds to a lag behind the Sun of $\sim
100$~km/s. This is dissimilar to the value for the canonical thick
disk and to the value for the canonical stellar halo.  We selected
stars in restricted ranges of color and magnitude, in both $\ell =
270^\circ$ and $\ell = 90^\circ$, so that modulo metallicities, the
stars are typically at a few kiloparsecs from the Sun, and one can
infer that we have found a population of intermediate angular
momentum.  This population is detected  in a wide range of
Galactic latitude and longitude. Indeed, the K-giant sample of MFF
also shows indications of this population, particularly for [Fe/H]
$\simlt -1.5$ (see their Figure~7 (d) and (g)). As noted in the
introduction, this mean rotational velocity is what one would
expect for debris from a typical shredded satellite.  The thick disk
may well have been caused by the most massive merger suffered by the
Milky Way since the onset of star formation in the thin disk.  In this
case, the debris from the satellite implicated in this `minor merger'
would dominate over other debris. We propose that this is indeed the
case, and interpret our observations as tracing the remains of the
satellite whose merger with the Milky Way created the thick disk.  We
intend to make detailed comparisons with simulations of the merger
process to constrain further this last significant merger of the Milky
Way. 

\acknowledgments

RFGW thanks the Aspen Center for Physics for hospitality.  MIW acknowledges PPARC.    AK and EKG
acknowledge the Swiss NSF (grant 200020-105260). 
We thank the referee for his quick and extremely useful report.

\begin{deluxetable}{cccc}
\normalsize
\tablecaption{\large {Derived mean orbital rotational velocity}
\label{vrot.tab}}
\tablewidth{0pt}
\tablehead{
\colhead{Population  } & \colhead{$<V_\phi>$~km/s} & \colhead {$\sigma_{V\phi}$~km/s} & \colhead{Ref.} \\
}
\startdata
(Old) thin disk  & $\sim 210$ & 20 & 1 \\
Canonical thick disk & $\sim 175$ & 40 & 2  \\
stellar halo & $\sim 0$ & 100 & 3 \\
debris?  & $\sim 100$ & 40 & 4 \\
\enddata
\tablecomments{Refs: 1 - Dehnen \& Binney 1998 ; 2 - MFF; 3 - Norris 1994 ; 4 - GWN and this paper.}

\end{deluxetable}

\clearpage

\begin{figure}
\includegraphics[angle=270,scale=.65]{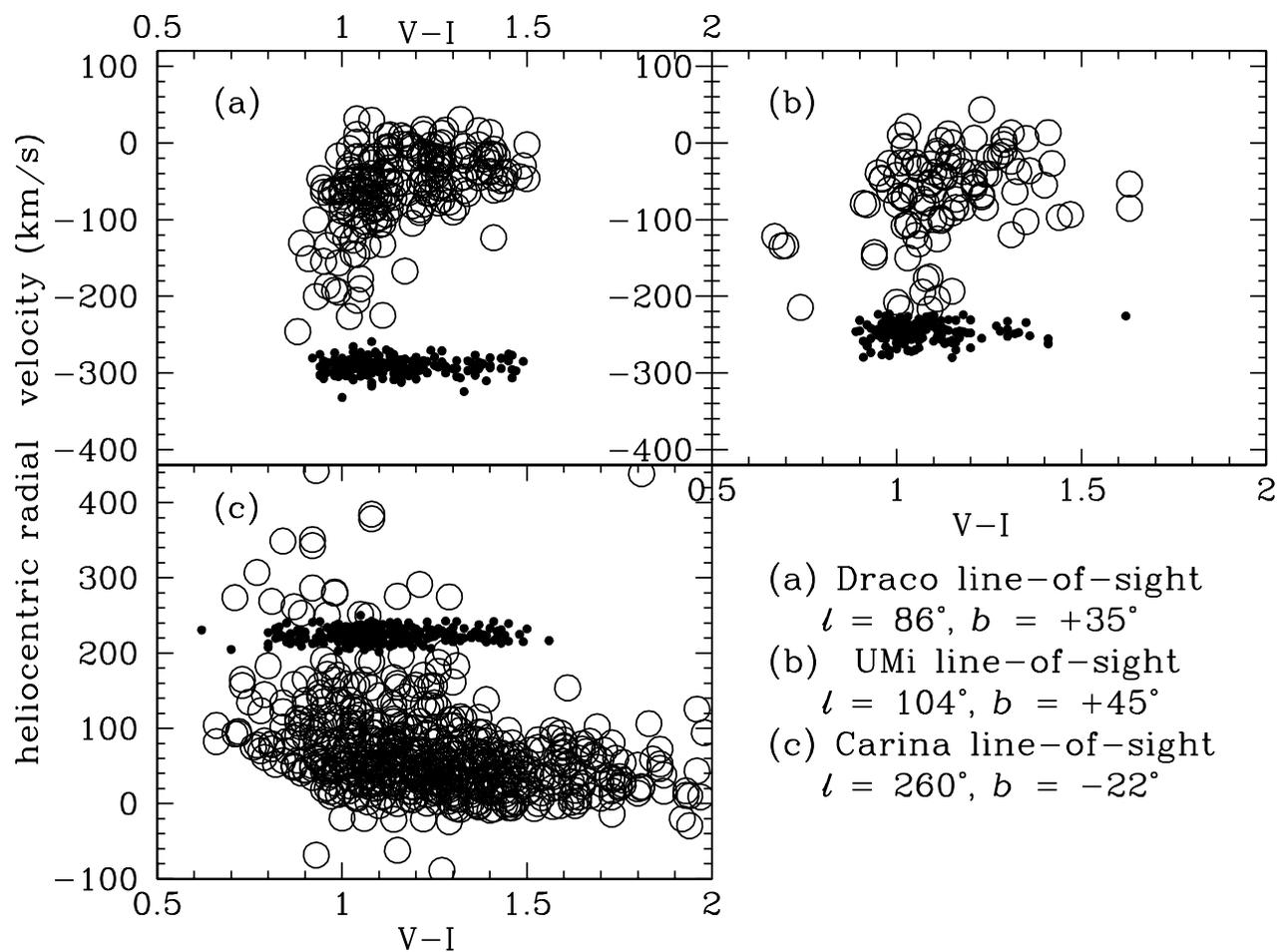}
\caption{Scatter plot of V$-$I color versus radial velocity for all
stars observed in the dwarf spheroidal lines-of-sight.  Stars with
very high probability to be members are indicated by small filled
circles, the likely non-members are indicated by large open circles. Typical errors are a few km/s, less than the symbol size.} 
\label{fig:fig1}
\end{figure}
\clearpage

\begin{figure}
\includegraphics[angle=270,scale=.65]{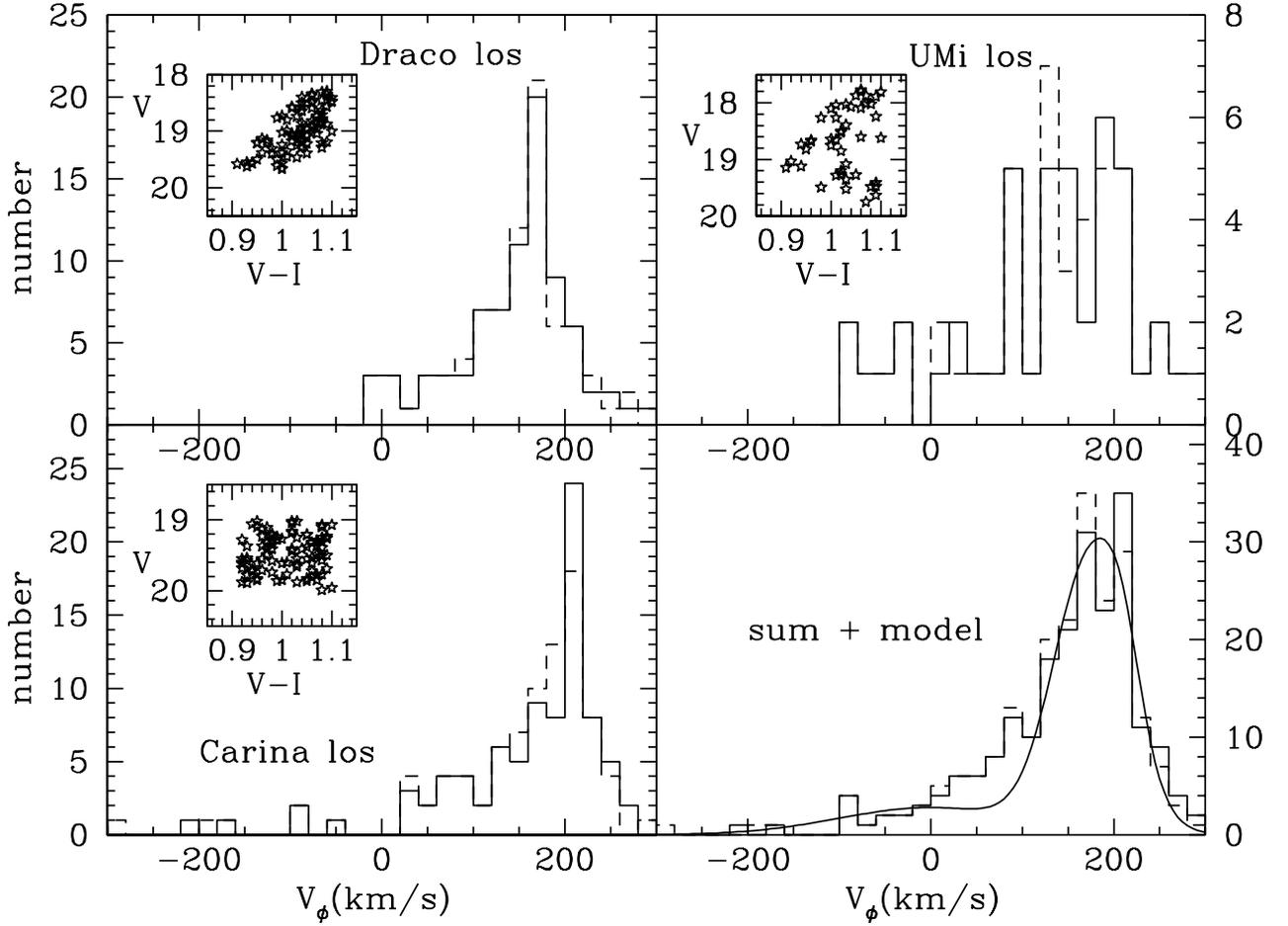}
\caption{Histograms of derived azimuthal streaming velocity,
$V_\phi = V_{Gal}/\beta$, for `G-stars' in the three lines-of-sight to 
dSph, plus their $V, \, V-I$ distributions.   The solid histograms adopt the isochrone with ${\rm [Fe/H]}
\sim -0.5$, while the dashed histograms adopt the more metal-poor
isochrone, ${\rm [Fe/H} \sim -1.5$.  The smooth curve in the lower right panel 
shows the expected kinematics for standard Galaxy models. 
The stars with derived  $V_\phi \sim 100$~km/s are not 
expected.}
\label{fig:fig2}
\end{figure}
\clearpage

\begin{figure}
\includegraphics[angle=270,scale=.65]{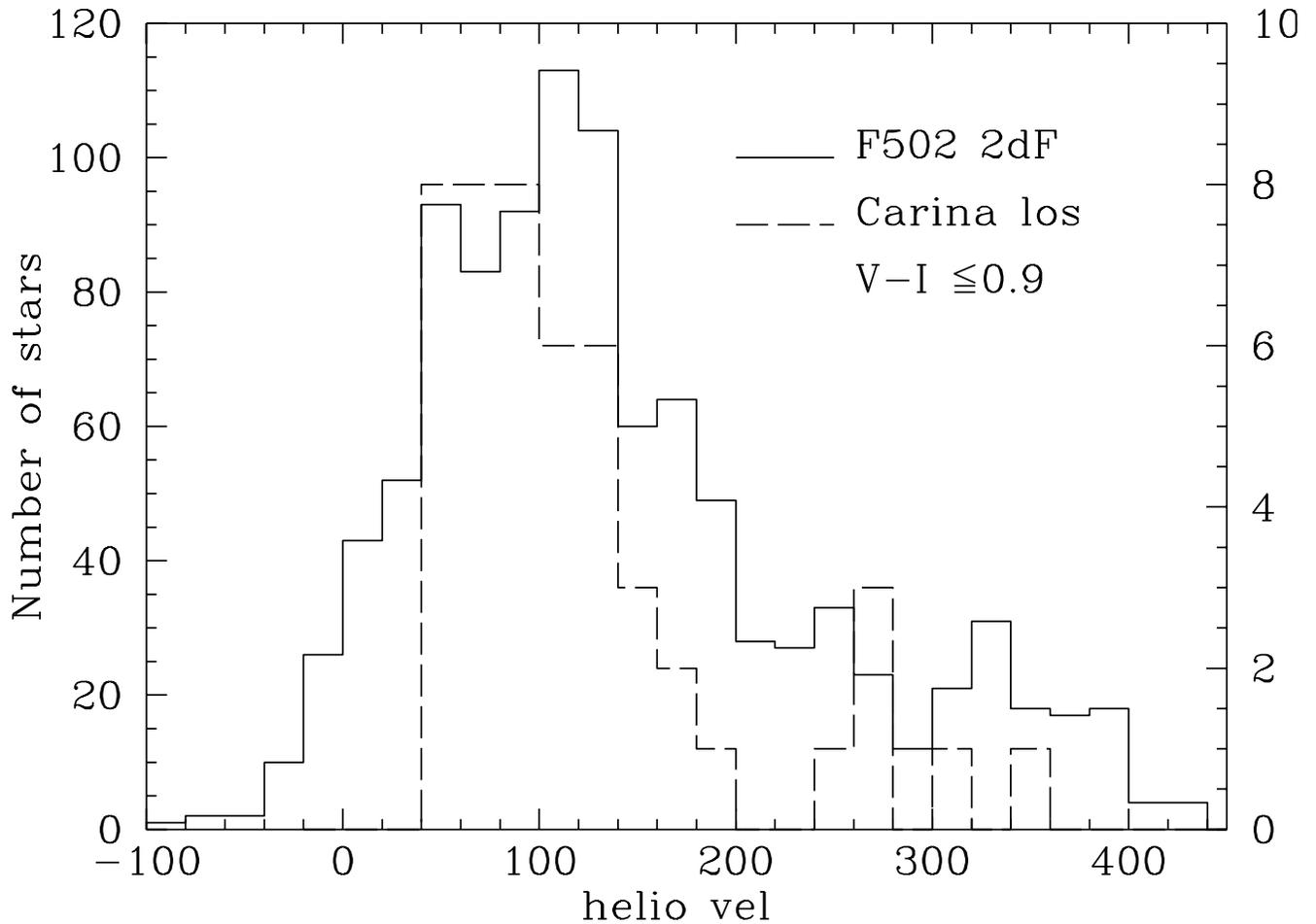}
\caption{Comparison between the heliocentric radial velocity
histograms for faint stars with $V \ge 18$  and that have F/G colors, in
the line-of-sight to the Carina dSph ($\ell \sim 260,\, b \sim -22$)
-- dashed histogram, and righthand y-axis --  and in field F502 ($\ell
\sim 270, \, b \sim +33$).  Given the difference in latitude, and the
resultant difference in projection of $V_\phi$ into the line-of-sight,
the agreement is very good.}
\label{fig:fig3}
\end{figure}

\clearpage

\end{document}